\documentclass[secnumarabic,aps,prb,amsfonts,twocolumn,%
showkeys,floatfix]{revtex4}

\usepackage{bm}

\newcommand{\gl}[1]{(\ref{#1})}

\begin{document}

\title{Superconductivity interpreted as $\vec k$-space magnetism}
\author{Ekkehard Kr\"uger}
\email{krueger@mf.mpg.de}
\affiliation{Max-Planck-Institut f\"ur Metallforschung, D-70506 Stuttgart,
  Germany}
\date{\today}
\begin{abstract}
  In preceding papers the author proposed a new mechanism of Cooper pair
  formation derived within an extended Heisenberg model. The new mechanism
  operates in narrow, partly filled ``superconducting'' energy bands of
  special symmetry and shows resemblances, and also great differences as
  compared with the familiar BCS mechanism. In the present paper the
  resulting superconducting state is interpreted as a state in which the
  spins are ordered within the $\vec k$ space. In this picture, the peculiar
  features of the new mechanism, as compared with the BCS mechanism, can be
  understood in a straightforward manner. On the one hand, the new mechanism
  resembles the BCS picture, because the formation of Cooper pairs is still
  mediated by bosons (having dominant phonon character in the isotropic
  lattices of the conventional superconductors). On the other hand, however,
  the pair formation is {\em not} the result of an {\em attractive}
  electron-electron interaction mediated by these bosons.  Rather, the bosons
  carry the crystal-spin angular momentum $1\cdot\hbar$ and generate in a
  new way {\em constraining forces} that constrain the electrons to form
  Cooper pairs. The scale of the transition temperatures in conventional and
  high-temperature superconductors is set by the excitation energies of {\em
    stable} crystal-spin-1 bosons that are different in isotropic and
  anisotropic materials.
\end{abstract}
\maketitle
\section{Introduction}
\label{introduction}
In previous papers\cite{es,josn} a new mechanism of Cooper pair formation was
proposed that operates in narrow partly filled ``superconducting'' energy
bands ($\sigma$ bands) of special symmetry. Except for their name, these
$\sigma$ bands have no resemblances or analogies with graphite-like $\sigma$
bands. The new mechanism can be derived within a nonadiabatic extension of
the Heisenberg model of magnetism,\cite{hei} the ``nonadiabatic Heisenberg
model'' (NHM) described in detail in Ref.~\onlinecite{enhm}. The NHM
emphasizes in a new way the atomiclike character of the electrons in narrow
bands as described by Mott\cite{mott} and Hubbard\cite{hubbard}: the
electrons occupy the localized states {\em as long as possible} and perform
their band motion by hopping from one atom to another. Within the NHM these
localized states are represented by (spin-dependent) Wannier functions which
form an {\em exactly} unitary transformation of the Bloch functions of the
considered energy bands.

The new mechanism of Cooper pair formation resembles the familiar mechanism
presented within the Bardeen-Cooper-Schrieffer (BCS) theory\cite{bcs} because
the formation of Cooper pairs is still mediated by bosons. In contrast to the
BCS mechanism, however, these bosons carry the crystal spin $1\cdot\hbar$
and must be sufficiently stable to transport it through the crystal. Further,
the formation of Cooper pairs is {\em not} the result of an {\em attractive}
electron-electron interaction but is constrained in a new way by quantum
mechanical constraining forces operating in $\sigma$ bands.

These constraining forces have been illustrated in Ref.~\onlinecite{josi} in
terms of ``spring-mounted'' Cooper pairs.  In the present paper we interpret
the superconducting state within a narrow $\sigma$ band as ``$\vec k$-space
magnetism''. This new form of magnetism complies with the conservation law of
spin angular momentum only if the electrons form Cooper pairs. Within this
picture, the essential features of the new mechanism of Cooper pair formation
as given in the preceding paragraph can be clearly understood.

A set of energy bands of a metal is called $\sigma$-band complex if the Bloch
functions belonging to this set can be unitarily transformed into
spin-dependent Wannier functions (spin-dependent Wfs) which are best
localized, symmetry adapted to the paramagnetic group of this metal and
situated at the atoms.\cite{enhm} A $\sigma$-band complex contains just as
many bands as there are atoms in the unit cell. However, only those atoms
must be considered which are responsible for the superconducting state.

For simplicity, in this paper we assume the $\sigma$-band complex to consist
of a single band. The spin-dependent Wfs of such a single $\sigma$ band are
shortly defined in Section \ref{sec:1} and, more detailed, in
Refs.~\onlinecite{josn} or \onlinecite{josm}. Within the NHM, the peculiar
properties of the spin-dependent Wfs lead to a special operator
$H_{Cb}^{n\sigma}$ of Coulomb interaction shortly described in Section
\ref{sec:2}. The interaction $H_{Cb}^{n\sigma}$ is proposed to be responsible
for superconductivity.\cite{es,josn}

In Section \ref{sec:3} $H_{Cb}^{n\sigma}$ is approximated by a purely
electronic operator $H_{Cb}^{st}$, and in Section \ref{sec:4} $H_{Cb}^{st}$
is interpreted as a magnetic interaction that produces a spin structure in
the $\vec k$ space. Within this spin structure, however, electronic
scattering processes violate the conservation of spin angular momentum. As
shown in the following Section \ref{sec:5}, the spin structure is stable if
the electrons form Cooper pairs coupled via crystal-spin-1 bosons.

\section{Spin-dependent Wannier functions}
\label{sec:1}
Consider a metal with one atom in the unit cell and assume that this metal
possesses a narrow, half-filled $\sigma$ band in its calculated band
structure.

The Bloch functions $\varphi_{\vec k}(\vec r)$ of this band can be unitarily
transformed into spin-dependent Wfs
\begin{equation}
w_m(\vec r - \vec T, t) = 
\frac{1}{\sqrt{N}}\sum_{\vec k}^{BZ}e^{-i\vec k\cdot \vec T} 
\phi_{\vec km}(\vec r,t)
\label{sdwf}
\end{equation}
with the functions    
\begin{equation}
\phi_{\vec km}(\vec r,t) = \sum_{s = -\frac{1}{2}}^{+\frac{1}{2}}
f_{sm}(\vec k)u_{s}(t)\varphi_{\vec k}(\vec r)
\label{sdbf}
\end{equation} 
being ``spin-dependent Bloch functions'' with spin directions depending on
the wave vector $\vec k$. The two-dimensional matrix
\begin{equation}
  \label{eq:6}
\bm{f}(\vec k) =  [f_{sm}(\vec k)]
  \end{equation}
  is, for each $\vec k$, unitary and
\begin{equation}\
 u_{s}(t) =
\delta_{st}
\label{paulisf}
\end{equation}
stands for Pauli's spin function with the spin quantum number $s = \pm
\frac{1}{2}$ and the spin coordinate $t = \pm \frac{1}{2}$. 

The quantum number $m = \pm\frac{1}{2}$ of the crystal spin distinguishes
between the two Wannier functions belonging to the same position $\vec T$.
If in Eq.~\gl{sdbf} we have
\begin{equation}
f_{sm}(\vec k) = \delta_{sm},
\label{deltasm}
\end{equation}
the two functions $\phi_{\vec km}(\vec r,t)$ with $m = \pm\frac{1}{2}$ are
usual Bloch functions with the spins lying in $+z$ and $-z$ direction,
respectively. Otherwise, these functions still are usual Bloch functions with
antiparallel spins that lie, however, in a direction $z'$ different from the
$z$ direction.

In a $\sigma$ band the matrix $\bm{f}(\vec k)$ in Eq.~\gl{eq:6} can be
chosen in such a way that 

(1) the spin-dependent Bloch functions $\phi_{\vec km}(\vec r,t)$ vary
smoothly through the whole $\vec k$ space, and 

(2) the $w_m(\vec r - \vec T, t)$ are adapted to the symmetry of the
considered metal. 

In particular, by application of the operator $K$ of time inversion we obtain
\begin{equation}
Kw_m(\vec r - \vec T, t) = \pm w_{-m}(\vec r - \vec T, t)
\label{C9}
\end{equation}
(where the plus is defined to belong to $m = +\frac{1}{2}$ and the minus to
$m = -\frac{1}{2}$). The smoothness of the spin-dependent Bloch functions
$\phi_{\vec km}(\vec r,t)$ guarantees that the spin-dependent Wfs are
optimally localizable.

It is essential that in a $\sigma$ band the matrix $\bm{f}(\vec k)$ is not
independent of $\vec k$ if the Wfs comply with the above conditions (1) and
(2).\cite{josn}

\section{Spin-boson interaction}
\label{sec:2}
The nonadiabatic Coulomb interaction 
\begin{widetext}
\begin{equation}
H_{Cb}^{n\sigma} = \sum_{\vec T, m}\langle\vec T_{1}'', l_{1}; 
\vec T_{2}'', l_{2};\vec T_{1}, m_{1}, n; 
\vec T_{2}, m_{2}, n|H_{Cb}|
\vec T_{1}', m_{1}', n; \vec T_{2}', m_{2}', n\rangle\,
b_{\vec T_{1}''l_{1}}^{\dagger}
b_{\vec T_{2}''l_{2}}^{\dagger}
c_{\vec T_{1}m_{1}}^{n\dagger}
c_{\vec T_{2}m_{2}}^{n\dagger}
c_{\vec T_{2}'m_{2}'}^{n}
c_{\vec T_{1}'m_{1}'}^{n} + \mbox{H.c.}
\label{sph}
\end{equation}
\end{widetext}
in a narrow half-filled $\sigma$ band depends also on boson operators
$b_{\vec Tl}^{\dagger}$ and $b_{\vec Tl}$ in order that it conserves the
total crystal spin,
\begin{equation}
[H_{Cb}^{n\sigma}, M(\alpha )] = 0 \mbox{ for } \alpha \in G_0,
\label{sphm}
\end{equation}
see Ref.~\onlinecite{josn}. The symmetry operators $M(\alpha )$ of the
crystal spin are defined for the electrons in Ref. \onlinecite{enhm} and for
the bosons in Ref.~\onlinecite{es}, and $G_0$ denotes the point group. The
fermion operators $c_{\vec Tm}^{n\dagger}$ and $c_{\vec Tm}^{n}$ create and
annihilate electrons with crystal spin $m$ in the nonadiabatic localized
states $|\vec T, m, n \rangle$ and the boson operators $b_{\vec
  Tl}^{\dagger}$ and $b_{\vec Tl}$ create and annihilate localized bosons
$|\vec T, l\rangle$ (with the crystal spin $l = -1, 0, +1$) which are
sufficiently stable to transport crystal spin angular momenta through the
crystal.

The nonadiabatic states $|\vec T, m, n \rangle$ are represented by
nonadiabatic localized functions of the form $\langle \vec r,t,\vec q~|\vec
T, m, n \rangle$, where the new coordinate $\vec q$ stands for that part of
the motion of the center of mass of the localized state $|\vec T, m, n
\rangle$ which nonadiabatically follows the motion of the electron occupying
this state.\cite{enhm} We may imagine that $\vec q$ denotes the {\em
  acceleration} of the center of mass.

\section{Static approximation within the nonadiabatic Heisenberg model}
\label{sec:3}
Assume the atomic cores in the considered crystal to be rigid and
absolutely immovable. On this fixed atomic array we approximate the operator
$H_{Cb}^{n\sigma}$ by the operator
\begin{eqnarray}
H_{Cb}^{st}  &=& \sum_{\vec T, m}\langle\vec T_{1}, m_{1}; 
\vec T_{2}, m_{2}|H_{Cb}|
\vec T_{1}', m_{1}'; \vec T_{2}', m_{2}'\rangle\nonumber\\*
&&\times c_{\vec T_{1}m_{1}}^{\dagger}
c_{\vec T_{2}m_{2}}^{\dagger}
c_{\vec T_{2}'m_{2}'}
c_{\vec T_{1}'m_{1}'}
\label{hcbn}
\end{eqnarray}
not depending on boson operators and, nevertheless, conserving the crystal
spin,
\begin{equation}
[H_{Cb}^{st}, M(\alpha )] = 0 \mbox{ for } \alpha \in G_0.
\label{sphs}
\end{equation}

The fermion operators $c_{\vec Tm}^{\dagger}$ and $c_{\vec Tm}$ create and
annihilate electrons in localized states $|\vec T, m\rangle$ which no longer
depend on $n$ and, hence, are represented by the spin-dependent Wfs $w_m(\vec
r - \vec T, t)$.

$H_{Cb}^{st}$ commutes with the space group operators because the $w_m(\vec r
- \vec T, t)$ are adapted to the symmetry of the considered metal.\cite{josn}
In particular, from Eq.~\gl{C9} it follows that $H_{Cb}^{st}$ commutes with
the operator $K$ of time inversion,
\begin{equation}
  \label{eq:3}
 [H_{Cb}^{st}, K] = 0.
\end{equation}
 
As a consequence of Eq.~\gl{sphs}, $H_{Cb}^{st}$ does not conserve the
electron spin $s$, see the following Section \ref{sec:4}. Therefore, this
``static approximation'' within the NHM differs from the familiar adiabatic
approximation. It is the natural approximation that ignores the motion of the
atomic cores when we start from the NHM.

\section{The static ground state within a narrow $\sigma$ band}
\label{sec:4}
Let ${\cal E}^{st}$ be the electron system of a narrow, half-filled $\sigma$
band represented by the static operator
\begin{equation}
\label{eq:1}
H^{st} = H_{HF} + H_{Cb}^{st}
\end{equation}
(where $H_{HF}$ stands for the Hartree-Fock operator). In ${\cal E}^{st}$ the
crystal spin $m$ of the spin-dependent Bloch functions $\phi_{\vec km}(\vec
r,t)$ given in Eq.~\gl{sdbf} is a conserved quantity. Consequently, at any
scattering process of two Bloch electrons in ${\cal E}^{st}$ the spin
directions of the scattered electrons are (slightly) changed. This leads to a
ground state $|G^{st}\rangle$ in which the Bloch states $|\vec km\rangle$
have $\vec k$-dependent spin directions as determined by the two-dimensional
unitary matrix $\bm{f}(\vec k)$ in Eq.~\gl{eq:6}. (It should be noted that
the crystal spin $m$ and the spin $s$ without the addition ``crystal'' are
different.  The former denotes the crystal spin of the nonadiabatic localized
states and the bosons, and the latter the spin of the naked electrons.)

The matrix $\bm{f}(\vec k)$ varies smoothly as a function of $\vec k$.  It
determines the symmetry and (sensitively) the localization of the
spin-dependent Wfs. Both their symmetry and their localization have an
important {\em physical meaning} within the NHM: The Wfs must be adapted to
the symmetry of the considered metal in order that the nonadiabatic
Hamiltonian commutes with the operators of the space group.  Further, the Wfs
must be optimally localized in order that the Coulomb energy in the $\sigma$
band is as low as possible. Strictly speaking, the Wfs are ``optimally''
localized if the energy difference $\Delta E$ in Eq.~(2.20) of
Ref.~\onlinecite{enhm} is maximum, see the detailed description of the NHM in
Ref.~\onlinecite{enhm}.

Hence, the directions $z'$ of the electron spins are fixed in
$|G^{st}\rangle$ by the condition that the Coulomb energy is minimum.
Therefore, we interpret the interaction $H_{Cb}^{st}$ as ``magnetic''
spin-spin interaction that produces Bloch states with $\vec k$-dependent spin
directions $z'$. However, the state $|G^{st}\rangle$ differs to some aspects
from a magnetic state in the local space: First, $|G^{st}\rangle$ is
invariant under time inversion,
\begin{equation}
  \label{eq:2}
K|G^{st}\rangle = |G^{st}\rangle,  
\end{equation}
since $H^{st}$ commutes with $K$. (Any magnetic order in the local space is
not invariant under time inversion. Within the NHM, such a state exists only
if the nonadiabatic Hamiltonian $H^n$ does not commute with $K$,
too.\cite{ef}) Second, the spins are not ordered in the sense that the spins
of the Bloch states have only one direction, say $+z'$ direction, but the
spins may either lie in $+z'$ or in $-z'$ direction with only the $z'$
direction being fixed.  Therefore, this new magnetic interaction generating
the $\vec k$-dependent $z'$ direction of the spins cannot be calculated by
exchange integrals and, hence, should not be called ``exchange interaction''.
Thirdly, the pure magnetic state $|G^{st}\rangle$ is not realized in nature,
see the following Section \ref{sec:5}.

\section{Cooper pair formation in a narrow $\sigma$ band}
\label{sec:5}
Starting from the magnetic state $|G^{st}\rangle$ we may understand in a
straightforward way the formation of Cooper pairs within the NHM.

At any scattering process in the electron system ${\cal E}^{st}$ the total
electron spin of the scattered electrons is not conserved. Hence, the
electrons must interchange spin angular momenta with the lattice of the
atomic cores. Such a mechanism produces deformed and accelerated atomic cores
and cannot be understood within the static approximation. Rather it is
described by the nonadiabatic Coulomb interaction $H_{Cb}^{n\sigma}$
representing a system in which at any electronic scattering process two
crystal-spin-1 bosons are excited or absorbed.

We may assume that at zero temperature the crystal-spin-1 bosons are only
virtually excited. That means that each boson pair is reabsorbed immediately
after its generation. Hence, whenever a boson pair is excited during a
certain scattering process 
\begin{equation}
  \label{eq:7}
\vec k_1', \vec k_2' \rightarrow\vec k_1, \vec k_2  
\end{equation}
of two electrons, this boson pair must be reabsorbed immediately after its
generation during a second scattering process
\begin{equation}
  \label{eq:8}
\vec k_3', \vec k_4' \rightarrow\vec k_3, \vec k_4  
\end{equation} 
of two other electrons.

Such a state can be studied within ${\cal E}^{st}$. Here we must look for
scattering processes
\begin{equation}
  \label{eq:9}
\vec k_1', \vec k_2',\vec k_3', \vec k_4' \rightarrow\vec k_1, \vec k_2,\vec
k_3, \vec k_4 
\end{equation}
conserving the total electron spin. Only in this case, the boson pair created
during the first process \gl{eq:7} is reabsorbed during the
second process \gl{eq:8}. 

At first sight, such scattering processes seem not to exist in ${\cal
  E}^{st}$ because the matrix $\bm{f}(\vec k)$ cannot be chosen independent
of $\vec k$ in a $\sigma$ band. However, from Eq.~\gl{C9} the equation
\begin{equation}
  \label{eq:4}
f_{sm}(-\vec k) = \pm f_{-s,-m}^*(\vec k)
\end{equation}
may be derived\cite{josn} showing that the spins in the Bloch state $|\vec
km\rangle$ and in its time-inverted state $|-\vec k-m\rangle$ lie exactly
opposite. 

Thus,\cite{josn} in ${\cal E}^{st}$ we can construct (symmetrized) Cooper
pairs
\begin{equation}
  \label{eq:5}
\beta^{\dagger}_{\vec k} = c_{\vec k\uparrow}^{\dagger}c_{-\vec
  k\downarrow}^{\dagger} - c_{\vec k\downarrow}^{\dagger}c_{-\vec
  k\uparrow}^{\dagger}. 
\end{equation}
with (exactly) zero total spin, where the fermion operators $c_{\vec
  ks}^{\dagger}$ create Bloch electrons with spin $s =\,
\uparrow,\downarrow$. Consequently, scattering processes of the form
\begin{equation}
  \label{eq:10}
  \vec k', -\vec k' \rightarrow\vec k, -\vec k
\end{equation}
conserve the total spin angular momentum within ${\cal E}^{st}$.  

We now may construct the nonadiabatic state in a narrow half-filled
$\sigma$ band in which any boson pair is reabsorbed immediately
after its creation.

Assume all the $N$ electrons (with $N$ being even) in ${\cal E}^{st}$ to form
Cooper pairs. Consequently, the total electron spin in ${\cal E}^{st}$ is
exactly zero. Two of these Cooper pairs we denote by
\begin{equation}
  \label{eq:11}
(\vec k_1', -\vec k_1')\ \mbox{ and }\ (\vec k_2', -\vec k_2').   
\end{equation}
Assume further the two Bloch electrons with wave vectors $\vec k_1'$ and
$\vec k_2'$ to be scattered into the states $\vec k_1$ and $\vec k_2$,
\begin{equation}
  \label{eq:12}
\vec k_1', \vec k_2' \rightarrow\vec k_1, \vec k_2.  
\end{equation}
After this process the total spin angular momentum in ${\cal E}^{st}$ is no
longer zero since the two Cooper pairs \gl{eq:11} are destroyed. Hence, a
boson pair was emitted during this scattering process.  It may be reabsorbed
during a subsequent scattering process if then the total spin in ${\cal
  E}^{st}$ again is zero, that means, if then all the electrons in ${\cal
  E}^{st}$ again form Cooper pairs. Thus, at zero temperature, the process
\begin{equation}
\nonumber
-\vec k_1', -\vec k_2' \rightarrow-\vec k_1, -\vec k_2
\end{equation}
follows immediately upon the first process \gl{eq:12}. 

\section{Conclusions}
\label{sec:6}

The picture of superconductivity as $\vec k$-space magnetism presented in
this paper shows clearly the peculiar feature of the Cooper pair formation
within the NHM: Within the NHM any {\em attractive} electron-electron
interaction is unimportant for the formation of Cooper pairs. The bosons
(belonging to the interaction $H_{Cb}^{n\sigma}$) need not effect an
attractive interaction between the electrons, but only produce {\em
  constraining forces} that constrain the electrons to form Cooper pairs.

As in Ref.~\onlinecite{josi}, the Cooper pairs generated by these
constraining forces in a narrow $\sigma$ band can be illustrated in terms of
``spring-mounted'' Cooper pairs, cf. Fig.~3 in Ref.~\onlinecite{josi}: Let
${\cal P}^0$ be the subspace of the Hilbert space spanned by the $N$-electron
states in the $\sigma$ band in which all the electrons form Cooper pairs, and
assume all the electrons to be in ${\cal P}^0$. Whenever two electrons are
scattered out of ${\cal P}^0$, a boson pair is excited which can only be
reabsorbed when the electrons are scattered in such a way that again they lie
in ${\cal P}^0$.  Hence, the bosons behave like ``springs'' that push the
electrons back into ${\cal P}^0$.

The bosons that stabilize the Cooper pairs are, in any material, the
energetically lowest boson excitations of the crystal possessing the crystal
spin $1\cdot\hbar$ and being sufficiently stable to transport it through
the crystal. These ``crystal-spin-1'' bosons are localized excitations $|\vec
T,l\rangle$ of well-defined symmetry\cite{es} which move as Bloch waves
through the crystal. Most likely, these $|\vec T,l\rangle$ are coupled
phonon-plasmon modes which in the isotropic lattices of the standard
superconductors have dominant phonon character.\cite{ehtc} However, in the
one- or two-dimensional sublattices of the high-$T_c$ materials, phonons are
not able to carry crystal-spin angular momenta.\cite{ehtc} Here, the $|\vec
T,l\rangle$ are energetically higher lying crystal-spin-1 excitations of
dominant plasmon character leading within the BCS theory to higher transition
temperatures.

I suppose that only quantum mechanical constraining forces as described in
this paper are able to produce stable Cooper pairs. This supposition is
corroborated by both theoretical arguments\cite{josn} and calculated band
structures. I have already identified $\sigma$ bands in the band structures
of a great number of superconductors, while I could not find $\sigma$ bands
in the band structures of metals not becoming superconducting.\cite{es,es2}
In particular, on the basis of this new mechanism a theoretical
interpretation of the Matthias rule was possible in the framework of the BCS
theory.\cite{josm}

For an examination of calculated band structures in terms of
magnetic\cite{ef} or superconducting bands the symmetry notations of the
Bloch functions in the symmetry points of the Brillouin zone must be known.
Unfortunately, these symmetry notations are omitted in nearly all the
published band-structure calculations of the new superconductors.

\acknowledgements
{I thank Ernst Helmut Brandt for critical comments on the manuscript.}

\end{document}